\documentclass[acmsmall,screen]{acmart}
\AtBeginDocument{
  }
\usepackage{diagbox}
\usepackage{multirow}
\usepackage{setspace}
\usepackage{algorithm}
\usepackage{algorithmic}
\usepackage{pifont}
\usepackage{booktabs}
\usepackage{enumitem}
\usepackage{wrapfig}
\usepackage{graphicx}
\usepackage{fancyvrb}
\usepackage{textcomp}
\usepackage{colortbl}
\usepackage{color}
\usepackage{tabularx}
\usepackage{tabulary}
\usepackage{caption}
\usepackage{subcaption}
\usepackage{rotating}
\usepackage{wrapfig}
\usepackage{threeparttable}
\usepackage[misc]{ifsym}
\usepackage{array}
\usepackage{microtype}
\usepackage{xspace}
\usepackage{amsmath} 
\setcopyright{acmlicensed}

\begin{document}
\pagestyle{standardpagestyle}
\title{Characterize LSM-tree Compaction Performance via On-Device LLM Inference}

\author{Jiabiao Ding}
\email{ajaxs8319@gmail.com}
\orcid{}
\affiliation{
  \institution{School of Informatics, Xiamen University}
  \country{China}
}
\author{Yina Lv}
\email{elainelv95@gmail.com}
\orcid{0000-0003-3971-3123}
\affiliation{
  \institution{School of Informatics, Xiamen University}
  \country{China}
}
\author{Qiao Li}
\email{qiaoli045@gmail.com}
\orcid{0000-0002-4579-4268}
\affiliation{
  \institution{Department of Computer Science, Mohamed bin Zayed University of Artificial Intelligence, Abu Dhabi}\country{UAE}
}
\author{Zhirong Shen}
\email{zhirong.shen2601@gmail.com}
\orcid{0000-0003-2673-5868}
\affiliation{
  \institution{School of Informatics, Xiamen University}
  \country{China}
}
\author{Chun Jason Xue}
\email{Jason.xue@mbzuai.ac.ae}
\orcid{0000-0002-6431-9868}
\affiliation{
  \institution{Department of Computer Science, Mohamed bin Zayed University of Artificial Intelligence, Abu Dhabi}\country{UAE}
}
\thanks{
Yina Lv is the corresponding author.
}

\begin{abstract}
Modern key-value storage engines built on Log-Structured Merge-trees (LSM-trees), such as RocksDB and LevelDB, rely heavily on the performance of their compaction operations, which are impacted by a complex set of interdependent configuration parameters.
Manually tuning these parameters for optimal performance demands considerable expertise, while traditional auto-tuning approaches struggle with the enormous search space and low sample efficiency inherent to this domain.
In recent years, Large Language Models (LLMs) have demonstrated strong capabilities in code generation and logical reasoning, offering new possibilities for system optimization.
However, applying LLMs to \emph{real-time compaction tuning} in such latency-sensitive environments is a double-edged sword.
While large-scale LLMs can offer superior reasoning for strategy generation, their high inference latency and computational cost make them impractical for interactive, low-latency tuning.
In contrast, small-scale LLMs achieve low latency but often at the expense of reasoning accuracy and tuning effectiveness.
In this paper, we first evaluate this trade-off by analyzing the compaction-tuning performance and inference latency of LLMs at different scales in an LSM-tree-based tuning case.
We then characterize the performance of LSM-tree on RocksDB v8.8.1, with a focus on adjusting the key compaction-related parameters under db\_bench workloads.
Our experimental results show a clear positive correlation between model capability and tuning effectiveness.
\end{abstract}



\maketitle

\section{Introduction}\label{sec:introduction}
The Log‑Structured Merge‑tree (LSM‑tree) has been adopted by widely‑used storage systems including RocksDB \cite{RocksDB}, LevelDB \cite{LevelDB}, Cassandra \cite{Cassandra}, and HBase \cite{HBase}.
By organizing data into multiple levels and leveraging sequential I/O alongside background compaction, LSM-trees achieve high write throughput, making them well-suited for write-intensive workloads \cite{ATC2017TRIAD,2022Dremel,2024vLSM,ATC2019SILK,ATC2022vigilkv,HotStorage2024ELMo-Tune,ELMo-Tune-V2}.
Due to the nature of the LSM-tree, its performance is largely determined by the efficiency of its background compaction.
This continuous data reorganization process dictates critical trade-offs among write amplification, point and range read latency, and long-term space efficiency.
Compaction itself, however, introduces significant performance bottlenecks: overly aggressive or poorly timed compaction can degrade foreground write throughput through I/O contention, while insufficient compaction increases read latency as queries traverse more data levels.

To mitigate these issues, several works proposed to optimize the compaction process and mitigate its associated performance bottlenecks.
Early efforts focused on refining compaction policies, such as tiered strategies for separating hot and cold data \cite{ATC2017TRIAD} and coordination with the underlying file system to reduce I/O interference \cite{ATC2019SILK}. 
More recent works integrate learning to guide compaction decisions, such as cost-benefit prediction for task selection \cite{2022Dremel} and fine-grained I/O scheduling based on read latency impact \cite{ATC2022vigilkv}.
At the architectural level, virtual compaction layers have been introduced to enable dynamic policy switching \cite{2024vLSM}.
These optimizations collectively aim to improve the efficiency and adaptability of compaction, yet they often operate within fixed or manually tuned parameter configurations, leaving the challenge of real-time, workload-aware parameter adaptation largely unaddressed.
Several works have attempted to mitigate these bottlenecks by tuning a large set of configuration parameters, often exceeding one hundred in number, that control memory allocation, I/O scheduling, compaction policies, and thread management.
These parameters exhibit strong interdependencies, making manual tuning labor-intensive and heavily reliant on expert knowledge.
Auto-tuning approaches, including heuristic-based methods, Bayesian optimization \cite{EuroMLSys2021Bayesian}, and reinforcement learning \cite{SIGMOD2023learning}, support real-time parameter tuning.
However, these methods face significant limitations.
The vast configuration space results in poor sample efficiency, while dynamic workloads and varying hardware conditions demand continuous adaptation, further complicating the search for optimal configurations.
Moreover, many existing approaches lack the ability to generalize across different operational environments, often requiring retraining or recalibration when deployment conditions change, which restricts their practicality in real-world, dynamic systems.

Large Language Models (LLMs) have been applied in various systems to leverage the remarkable capabilities in code generation, natural‑language reasoning, and even decision‑making in structured domains.
These abilities suggest that LLMs could be leveraged to interpret runtime metrics, understand workload patterns, and recommend parameter adjustments in a human‑like manner.
However, deploying LLMs for real-time or near-real-time parameter tuning introduces a performance–accuracy trade-off that is critical in latency-sensitive environments.
On one hand, large-scale models, often comprising hundreds of billions of parameters, provide superior reasoning accuracy, broader contextual understanding, and stronger generalization from limited examples.
These attributes make them well-suited for generating sophisticated tuning strategies.
However, their high inference latency and substantial computational and memory requirements typically necessitate execution on powerful, often cloud-based, hardware.
This introduces communication overhead and makes them impractical for integration into lightweight, low-latency control loops where decisions must be made within milliseconds.
On the other hand, lightweight on-device models, typically under 10 billion parameters, are designed for efficient execution on constrained hardware.
They offer fast inference times and low resource overhead, making them feasible for embedding directly within the storage engine to support real-time decision-making.
However, this gain in responsiveness often comes at the expense of reduced reasoning accuracy, limited contextual capacity, and narrower task generalization.
These constraints can compromise the quality of the generated tuning decisions, potentially leading to suboptimal configurations that fail to adapt effectively to complex or shifting workload patterns.
Therefore, the challenge lies in balancing the requirements for low-latency, on-device execution with the demand for sufficiently accurate and reliable reasoning to drive effective real-time tuning.

To address these challenges, we first analyze the performance during compaction and the inference time cost of different-scale LLMs in the context of LSM-tree parameter tuning.
We focus on optimizing compaction performance by tuning key concurrency parameters as listed in Table \ref{tab:compactionParameters}, as these directly determine thread allocation and resource utilization during LSM‑tree background operations.
The experiments are conducted on RocksDB v8.8.1 under db\_bench workloads, showing a positive correlation between model capacity and tuning effectiveness.

The main contributions of this paper are as follows.
\begin{itemize}
\item We present the empirical study of the trade-off between inference latency and tuning effectiveness when using on-device LLMs for real-time LSM-tree compaction optimization.
Experiments across multiple LLM scales quantify how model size affects both reasoning accuracy and parameter tuning performance.
Large models tune effectively but exceed latency budgets, while small models meet real-time constraints but may obtain suboptimal configurations. 
\item We analyze the limitations of the small-scale LLM in inference reasoning. 
Our study motivates further research on lightweight, reasoning-capable models for LSM-tree optimization.
\end{itemize}

The rest of this paper is organized as follows. 
Section \ref{sec:background} reviews related work on LSM‑tree auto‑tuning and LLM‑based decision systems. 
Section \ref{sec:motivation} details our empirical analysis of LLMs for tuning tasks. 
Section \ref{sec:relatedwork} introduces the state-of-the-art work.
Finally, Section \ref{sec:conclusion} gives a conclusion.

\section{Background}\label{sec:background}

\subsection{Architecture of LSM-tree}
\begin{figure}[t]
    \centering
    \includegraphics[width=0.7\textwidth]{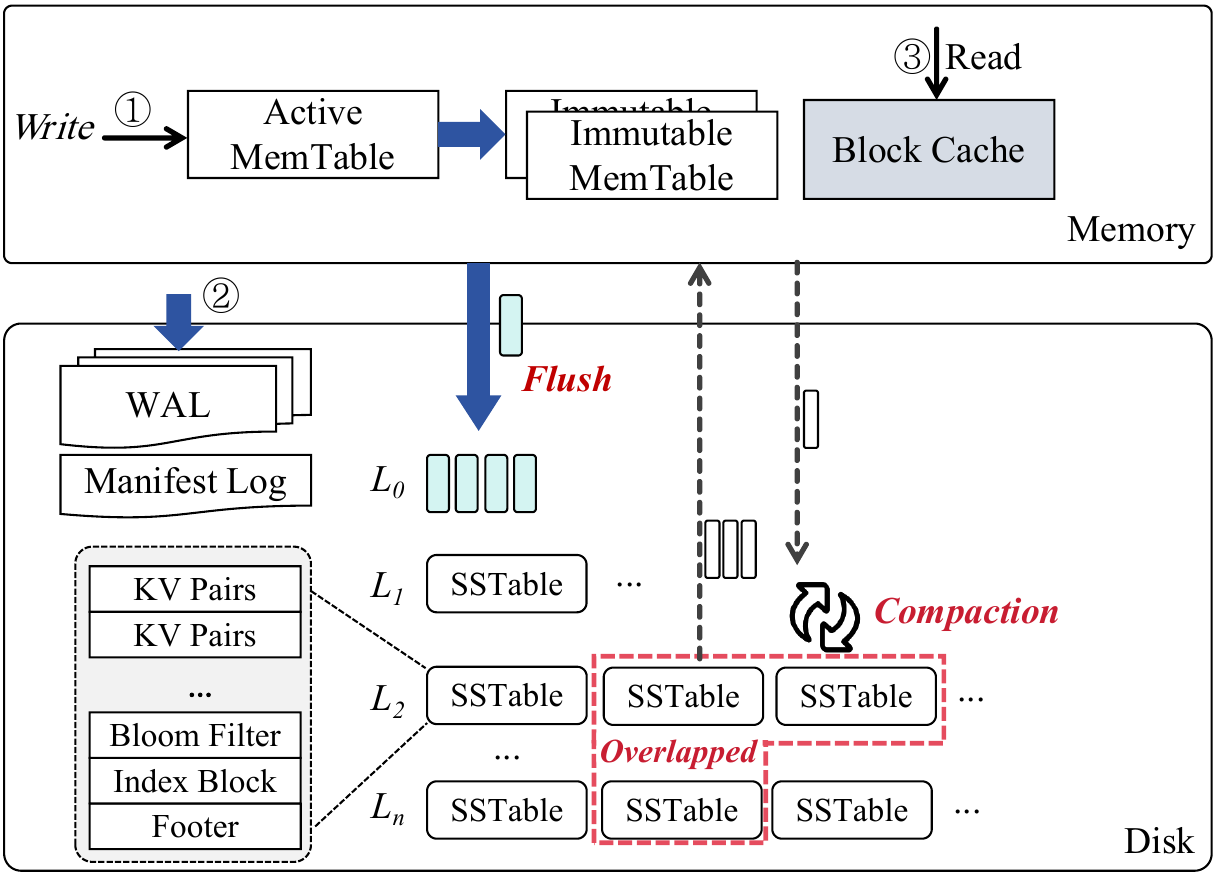}
    \caption{Architecture of a typical LSM-tree (e.g., RocksDB). \textit{Host writes are buffered in the MemTable, and marked as Immutable MemTable when full, and then flushed to immutable SSTables in $L_0$, and gradually compacted to larger, sorted SSTables in lower levels (i.e., $L_1$-$L_n$). Flush and compaction are managed by background threads.}}
    \label{fig:lsm_architecture}
\end{figure}

The Log-Structured Merge-tree (LSM-tree) is architected to optimize write performance through sequential I/O operations, while maintaining read efficiency via background data reorganization and indexing management.
Figure \ref{fig:lsm_architecture} illustrates the architecture of a typical LSM-tree (e.g., RocksDB \cite{RocksDB}).
As we can see, LSM-tree consists of two parts: an in-memory structure (MemTable) and multiple persistent levels on disk (SSTables).
This not only maximizes write throughput by rapidly absorbing writes into the MemTable but also supports progressive data compaction in the background and efficient point and range queries.
In the following, we introduce the detailed data management in such an LSM-tree.

\textbf{In-Memory Structure.}
When receiving a write operation (e.g., \texttt{put()}, \texttt{delete()}), the key-value pairs are initially buffered in an in‑memory structure known as the MemTable (Step \textcircled{1} in Figure \ref{fig:lsm_architecture}).
To guarantee data durability, each write is synchronously appended to a sequential Write‑Ahead Log (WAL) on persistent storage before an acknowledgment is returned to the host (Step \textcircled{2} in Figure \ref{fig:lsm_architecture}).
This WAL ensures that data can be reconstructed after a system failure, even if the MemTable has not yet been persisted.
The MemTable is commonly implemented as a skip list, balanced tree, or other ordered data structure that maintains keys in sorted order, thereby enabling efficient point lookups and range queries in memory.
The size of MemTable is determined by the configuration parameter \texttt{write\_buffer\_size}.
Once the data volume of the MemTable reaches this threshold, it transitions to an immutable state and is replaced by a newly allocated MemTable, allowing it to continue serving incoming writes.
This immutable MemTable is scheduled for a \textit{flush} operation, which asynchronously writes to disk in a sorted file.
The number of immutable MemTables is bounded by the parameter \texttt{max\_write\_buffer\_number}, which controls the maximum number of such pending flush tasks before write operations are stalled to manage memory pressure.

\textbf{On-Disk Storage Structure.}
When an immutable MemTable is flushed to disk, it is written into a Sorted String Table (SSTable).
As the immutable and sorted storage unit of the LSM-tree, this new SSTable is initially placed in Level 0 ($L_0$).
Each SSTable is organized into a sequence of data blocks (typically 4–64KB in size), which store the sorted key-value pairs.
To accelerate point queries, an index block is appended, mapping key ranges to the offsets of corresponding data blocks (Step \textcircled{3} in Figure \ref{fig:lsm_architecture}).
Additional metadata, such as Bloom filters and compression dictionaries, may also be included to further optimize read performance and storage efficiency.
Within $L_0$, SSTables may have \textit{overlapping key ranges} as they are flushed directly from memory in arbitrary order.
As a result, a point query in $L_0$ may need to examine every SSTable, potentially degrading read performance as the number of $L_0$ files grows.

To mitigate this issue, SSTables from Level 1 ($L_1$) onwards are enforced to have non-overlapping key ranges within the same level.
This organization is maintained by a background process called \textit{compaction}.
During compaction, selected SSTables from one level (e.g., $L_n$) are merged with overlapping SSTables from the next level ($L_{n+1}$), sorting the combined key set and rewriting it into new, larger SSTables in $L_{n+1}$.
This process not only eliminates key overlaps but also discards obsolete or deleted entries, thereby reducing space amplification and progressively consolidating data into fewer, more efficiently queryable files at deeper levels.
The size ratio between successive levels (e.g., \texttt{max\_bytes\_for\_level\_multiplier}) controls how aggressively data is merged and moved downward, directly influencing the trade-off between write amplification, read amplification, and storage overhead.

\textbf{Compaction Process.}
Compaction is used to reorganize on-disk data structures of the LSM-tree, including performance, space amplification, and read efficiency.
This is a background process that selects one or more SSTables from a level, merges them with overlapping SSTables from the next level, sorts the combined data, and writes compacted SSTables to the target level, while discarding obsolete or deleted entries.
The selection of SSTables is typically based on file size, file age, or overlap ratio.
This compaction process considers the following challenges.
First, it actively reclaims storage space by permanently eliminating stale or deleted entries, directly reducing space amplification.
Second, it enhances read performance by progressively reducing the number of SSTables that must be examined for point and range queries, thereby lowering read amplification and improving query latency.
Third, it implements a tiered data migration strategy that gradually moves data from smaller, performance-optimized levels (such as $L_0$ and $L_1$) to larger, capacity-optimized levels, effectively aligning data placement with storage hierarchy characteristics and access frequency patterns. 
The execution of compaction critically influences the trade-offs between write amplification (i.e., the total amount of physical writing relative to logical writing), read amplification (i.e., the number of disk reads per logical read), and space amplification (i.e., the ratio of physical storage used to logical data size).
There are two compaction strategies.
Leveled Compaction, where SSTables within each level (except $L_0$) maintain non-overlapping key ranges, typically merging one SSTable with all overlapping SSTables in the next level to minimize read amplification at the expense of higher write amplification; 
and Tiered Compaction, where SSTables are grouped into overlapping tiers within a level and merged in batches, reducing write amplification while increasing read amplification and temporary space overhead.

\subsection{Configurable Parameters}\label{sec:configParameter}
To adapt to various scenarios, LSM-tree-based storage engines provide an extensive set of configurable parameters, often exceeding one hundred in modern implementations such as RocksDB.
While parameters span memory management, caching, filtering, and I/O scheduling, those controlling compaction operations determine the critical performance during the triggering of background compaction operations.
Compaction, which performs continuous merging and reorganization of data across levels, directly determines the trade-offs among write amplification, read latency, and storage space efficiency. 
Its behavior is controlled by a group of tightly coupled parameters that collectively determine when, how, and with what resources compaction occurs.

\begin{table}[htbp]
\centering
\scriptsize
\caption{Key Compaction-Related Configuration Parameters and Default Values}
\label{tab:compactionParameters}
\begin{tabular}{@{}p{4cm}p{1.5cm}p{6.8cm}@{}}
\toprule
\textbf{Parameter} & \textbf{Default Value} & \textbf{Description and Impact on Compaction} \\
\midrule

\texttt{level0\_file\_num\_compaction\_trigger} & 4 & Number of $L_0$ files that triggers compaction to $L_1$. Higher values delay compaction but increase read amplification. \\

\texttt{level0\_slowdown\_writes\_trigger} & 20 & Number of $L_0$ files that triggers write slowdown. Helps control $L_0$ growth under heavy write workloads. \\

\texttt{level0\_stop\_writes\_trigger} & 36 & Number of $L_0$ files that completely stops writes until compaction reduces $L_0$ count. \\

\texttt{max\_bytes\_for\_level\_base} & 256 MB & Base size for $L_1$. Determines the size threshold for triggering compaction from $L_1$ to $L_2$. \\

\texttt{max\_bytes\_for\_level\_multiplier} & 10 & Multiplier between consecutive levels. Controls size growth across levels: $size(L) = base \times multiplier^{(L-1)}$. \\

\texttt{max\_background\_compactions} & 1 & Maximum number of concurrent background compaction threads. Increasing it speeds up compaction but raises CPU/I/O contention. \\

\texttt{max\_background\_jobs} & 2 & Total background jobs (compactions + flushes). Limits the total concurrency of background operations. \\

\texttt{target\_file\_size\_base} & 64 MB & Target SST file size for $L_1$. Affects file granularity and merge granularity during compaction. \\

\texttt{target\_file\_size\_multiplier} & 1 & Size multiplier for target file size per level. \\

\texttt{max\_compaction\_bytes} & / & Default value is \texttt{target\_file\_size\_base} $\times$ 25. Maximum bytes processed in a single compaction job. Limits compaction scope to control latency spikes. \\

\texttt{compaction\_readahead\_size} & 0 & Size of read-ahead during compaction. Non-zero values can improve compaction I/O performance on certain storage. \\

\texttt{compaction\_style} & \texttt{level} & Compaction algorithm: \texttt{level}, \texttt{universal}, or \texttt{fifo}. Determines compaction policy and shape of LSM-tree. \\

\texttt{compaction\_pri} & / & Priority policy for selecting files to compact: \texttt{kByCompensatedSize} (write amplification), \texttt{kOldestLargestSeqFirst} (time), or \texttt{kMinOverlappingRatio} (space amplification). \\

\bottomrule
\end{tabular}
\end{table}

The key compaction-related parameters can be categorized into three functional classes, as shown in Table \ref{tab:compactionParameters}. 
\\\textbf{(i) Concurrency Parameters:}
This category regulates the computational resources allocated to background compaction operations. The parameter \texttt{max\_background\_compactions} specifies the maximum number of concurrent compaction threads, directly controlling the degree of parallelism during merge operations. Its companion parameter \texttt{max\_background\_jobs} sets the upper bound for total background tasks, encompassing both compaction and flush operations. 
These parameters establish a critical trade-off: higher values increase merge throughput and reduce compaction backlog but compete with foreground operations for CPU cycles and I/O bandwidth. 
Insufficient concurrency leads to compaction lag, causing level imbalance and read degradation, while excessive concurrency can trigger resource contention that impairs both read and write latency.
The optimal setting depends on CPU core count, storage device characteristics, and workload concurrency patterns, requiring dynamic adjustment as system conditions evolve.
\\\textbf{(ii) Triggering Thresholds:}
These parameters define the precise conditions that initiate compaction activities, serving as the feedback mechanism between system state and maintenance operations.
The threshold pair \texttt{level0\_slowdown\_writes\_trigger} and \texttt{level0\_stop\_writes\_trigger} operates as a two-stage flow control mechanism: when $L_0$ accumulates files exceeding the slowdown threshold, incoming writes are deliberately delayed to allow compaction to catch up; if accumulation continues to the stop threshold, writes are paused entirely until compaction reduces the backlog.
This creates a delicate balance where aggressive thresholds minimize read amplification but increase write latency variability, while lenient thresholds improve write consistency at the cost of potential read performance degradation.
Similarly, level-based triggers like \texttt{max\_bytes\_for\_level\_multiplier} determine when upper levels become eligible for compaction based on size ratios.
These thresholds essentially implement the system's policy for trading immediate write latency against long-term read performance and space efficiency.
\\\textbf{(iii) Structure Parameters:} 
This class dictates the fundamental structural properties of the LSM-tree hierarchy, determining how data organizes itself across levels over time.
The base size parameter \texttt{max\_bytes\_for\_level\_base} establishes the capacity of Level 1, with subsequent levels growing geometrically by \texttt{max\_bytes\_for\_level\_multiplier}, collectively defining the tree's fanout and depth.
The companion parameter \texttt{target\_file\_size\_base} controls the granularity of SSTable files, influencing merge granularity and compaction selectivity.
These structural parameters have profound implications: a larger base size reduces write amplification by decreasing merge frequency, but increases read amplification due to wider key ranges per level; smaller file sizes improve compaction flexibility and hotspot adaptation but increase metadata overhead and reduce sequential I/O efficiency.
The geometric progression defined by these parameters essentially encodes the system's long-term strategy for distributing the cost of data organization across the lifetime of stored records.

These parameters exhibit strong interdependence and non-linear interaction.
For example, increasing compaction concurrency may reduce write amplification but can also contend with foreground I/O, degrading read and write latency if not balanced with appropriate memory and I/O rate limits.
Similarly, relaxing compaction triggers may improve short-term write throughput at the cost of accumulating more levels and degrading read performance over time.
This creates a dynamic optimization where manual tuning becomes infeasible and automated methods struggle to converge efficiently.

Existing automated tuning approaches, including heuristic search and reinforcement-learning, face significant challenges in this domain.
Reinforcement learning can learn adaptive policies, but it demands substantial training data and frequently fails to generalize across varying workloads and hardware configurations.
Moreover, these methods lack semantic awareness of the LSM-tree's internal logic.
They operate on numerical metrics without understanding the causal relationships between configuration choices and system outcomes, for instance, why a certain compaction strategy fails under a skewed read-write mix, or how thread allocation interacts with SSD parallelism.
This semantic gap limits their ability to reason interpretably, adapt to unseen states, or provide actionable insights beyond incremental tuning. 
The need for a tuning framework that combines low-latency decision-making with deeper semantic reasoning motivates our exploration of lightweight, on-device LLMs as a new paradigm for real-time compaction optimization.

\subsection{LLM Inference}
Large Language Model (LLM) inference offers a promising alternative for system tuning, with the potential to analyze runtime metrics, comprehend system states, and generate semantically meaningful tuning recommendations.
However, applying LLMs to real-time tuning scenarios introduces a critical trade-off. 
While models with larger parameter sizes (e.g., 256B) possess stronger reasoning capabilities and may deliver more accurate recommendations, their substantial computational requirements and high inference latency render them impractical for deployment alongside database instances that demand immediate responsiveness. 
In contrast, small-scale LLMs (e.g., \textless10B parameters) suitable for edge deployment offer fast inference and minimal resource consumption but may lack the reasoning fidelity required for complex tuning decisions.
\begin{figure}[htbp]
    \centering
    \includegraphics[width=0.9\textwidth]{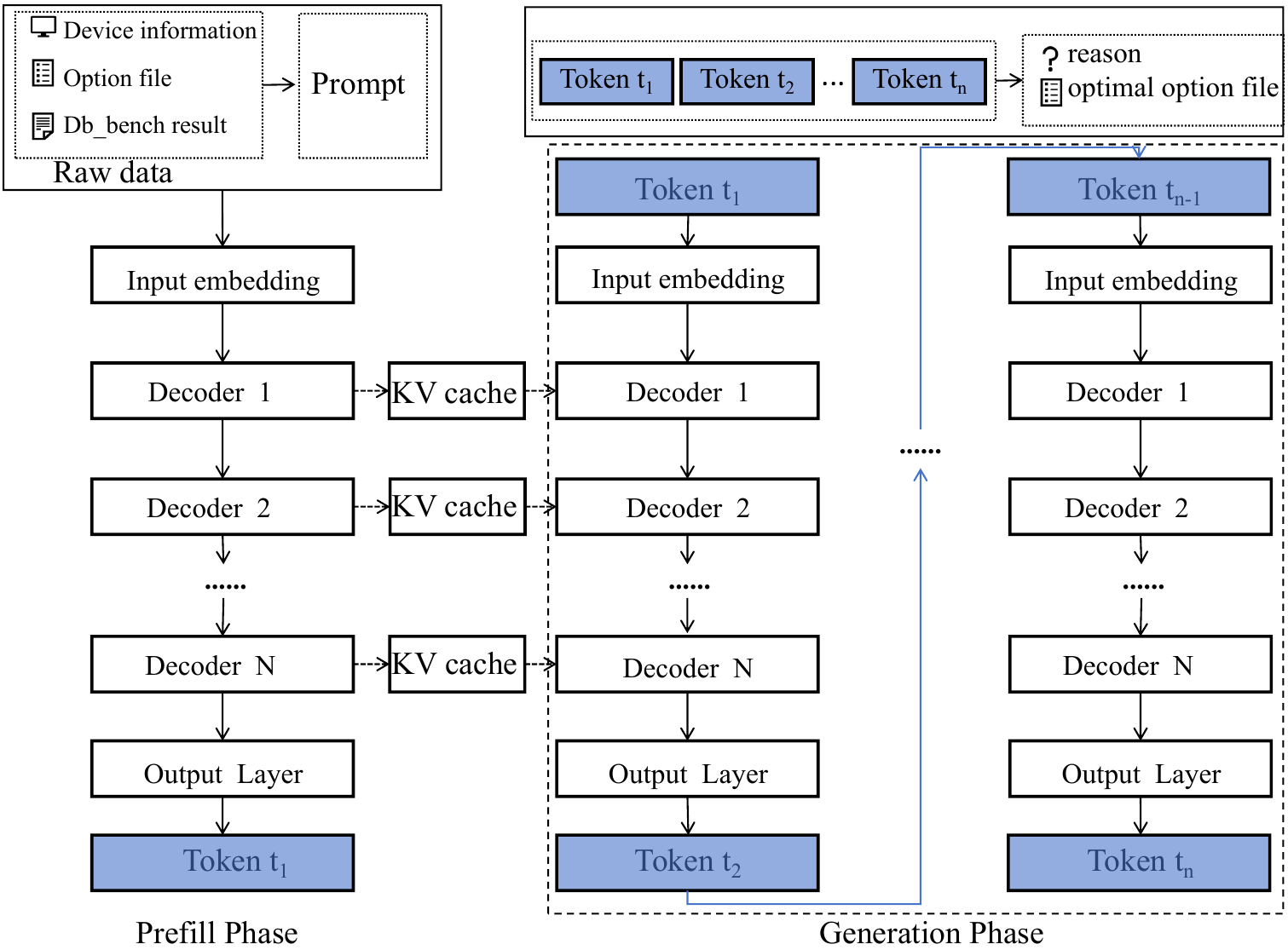}
    \caption{Workflow of LLM inference.}
    \label{fig:LLM}
\end{figure}

As illustrated in Figure \ref{fig:LLM}, the inference pipeline is designed to transform structured raw data, including device information, configuration files, and benchmark results, into optimized tuning decisions.
The process initiates with the Prompt Construction stage, where heterogeneous raw inputs are tokenized and structured into a coherent contextual prompt.
This carefully constructed prompt then guides the model to sequentially generate two outputs: first, a detailed diagnostic reasoning that interprets the current system state and performance patterns; second, a tailored optimized configuration file that translates this understanding into actionable parameter adjustments.

The LLM inference process consists of two phases: 
the Prefill Phase and the Autoregressive Generation Phase. 
During the prefill stage, the model processes the entire input prompt in parallel to compute initial embeddings and populate the Key-Value Cache (KV Cache).
Once the KV Cache is fully initialized and the first token ($t_1$) is generated, the system transitions to the autoregressive generation phase.
Here, $t_1$ serves as the initial input to generate the remaining tokens sequentially, one token at a time.
Each new token's prediction depends on the cumulative context stored in the KV Cache, which is updated incrementally after each generation step.
As illustrated, large-scale LLMs typically rely on a deep, multi-layer Transformer decoder stack (denoted as Decoder~1 through~N, where N can be dozens of layers).
For each generated token (e.g., from $t_1$ to $t_n$), the hidden states must traverse the entire depth of the network. 
While the KV Cache is utilized to store historical Key and Value states to avoid redundant attention computations, the massive size of LLM means that this cache grows linearly with sequence length, consuming a large amount of GPU memory and introducing memory bandwidth constraints.
In contrast, small-scale LLMs present a distinct advantage in this context. 
Due to their significantly reduced number of layers (smaller \(N\)) and hidden dimensions, small-scale LLMs significantly shorten the computational path for each forward pass. 
This architectural efficiency not only yields lower inference latency but also results in a much smaller KV Cache footprint. 
Such characteristics enable accelerated serial generation and support deployment on edge devices with limited VRAM capacity, standing in sharp contrast to the resource-intensive nature of their larger counterparts.

\section{Motivation}\label{sec:motivation}
LLMs present a promising alternative for system tuning by leveraging their pre-trained knowledge and causal reasoning capabilities to interpret performance metrics, discern workload patterns, and generate targeted tuning recommendations.
Prior studies have demonstrated the potential of cloud-sized large-scale LLMs, such as GPT-4 \cite{achiam2023gpt,baktash2023gpt} and DeepSeek-V3 \cite{DeepSeek}, in offline system optimization tasks. 
However, applying large-scale LLMs to real-time compaction tuning for edge-deployed LSM-trees introduces critical challenges that motivate the present work.
Existing LLM-based auto-tuning approaches typically rely on large-scale, cloud-hosted models with substantial parameter counts to ensure high inference accuracy. 
However, their considerable inference latency, recurring API costs, and dependence on stable network connectivity render them unsuitable for edge environments with strict resource and responsiveness constraints.

To validate this analysis, we conducted an empirical study on RocksDB and an LLM-driven tuning framework \cite{HotStorage2024ELMo-Tune} using the db\_bench tool. 
The target LSM‑tree engine is RocksDB v8.8.1, compiled using its default optimization flags.
Our experiment compares the tuning effectiveness and operational overhead of LLMs across multiple model scales, ranging from cloud-scale models (>200B parameters) to edge-suitable small-scale variants (\textless10B parameters).
This comparison shows the inherent trade-off between reasoning capacity and inference efficiency that is central to real-time tuning in resource-constrained environments. 
\begin{figure}[t]
    \centering
    \parbox{0.49\columnwidth}{
    \centering
    \includegraphics[width=1\linewidth]{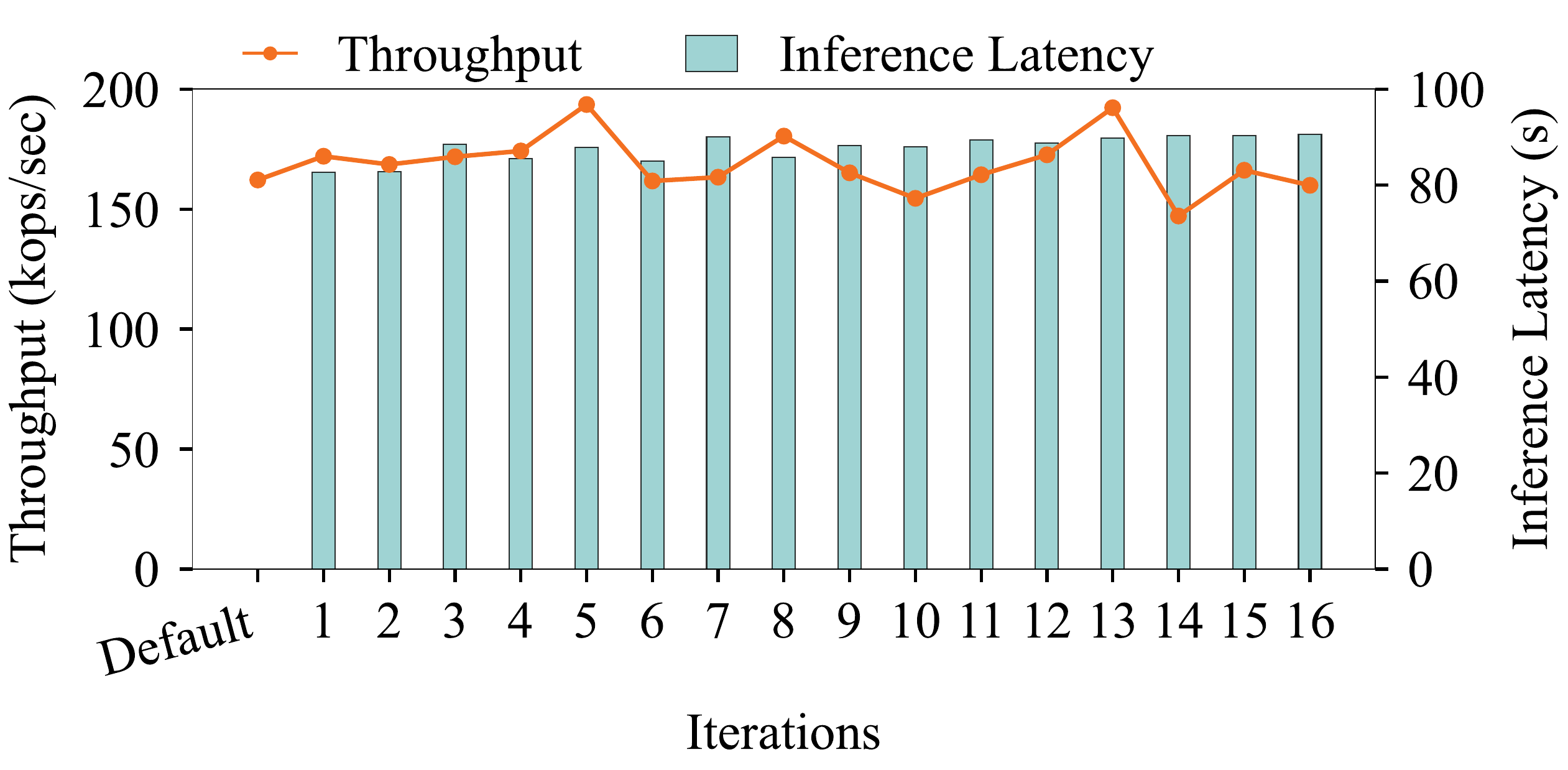}\\
    (a) \texttt{fillrandom (60s duration)}}\vspace{0.05in}
    \parbox{0.49\columnwidth}{
    \centering
    \includegraphics[width=1\linewidth]{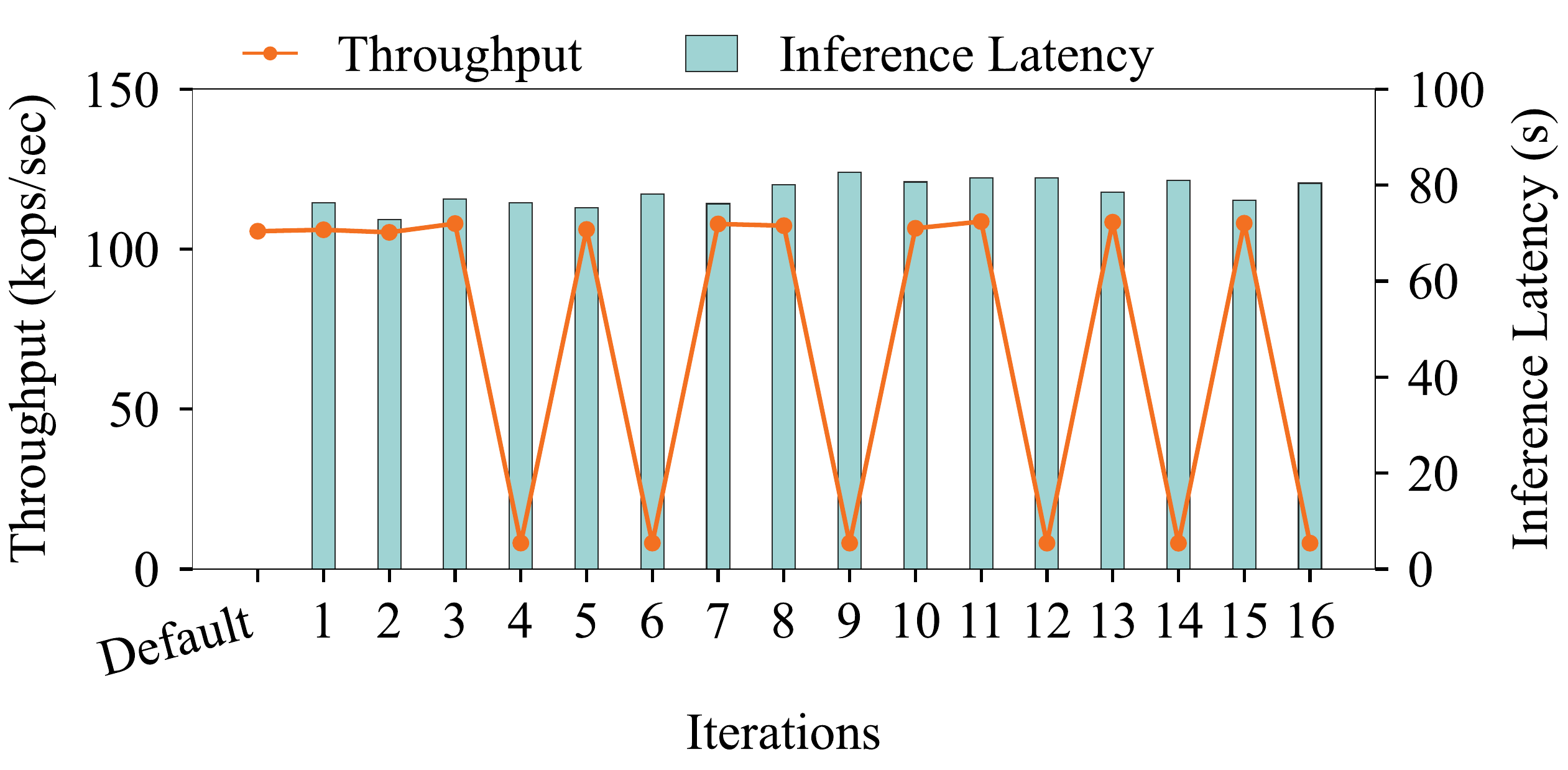}\\
    (b) \texttt{readrandom (60s duration)}}\vspace{0.05in}
    \parbox{0.49\columnwidth}{
    \centering
    \includegraphics[width=1\linewidth]{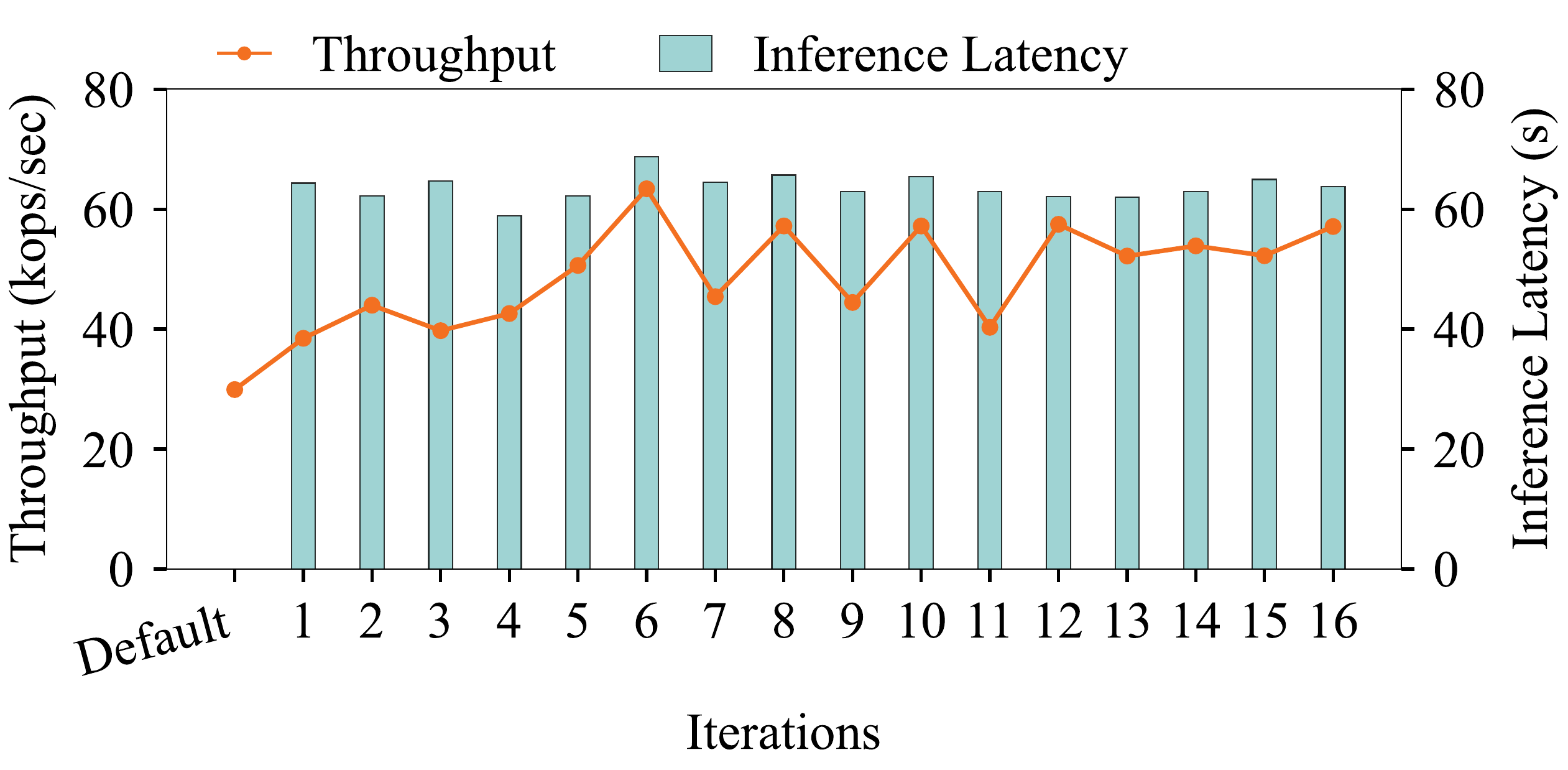}\\
    (c) \texttt{fillrandom (1000s duration)}}
    \parbox{0.49\columnwidth}{
    \centering
    \includegraphics[width=1\linewidth]{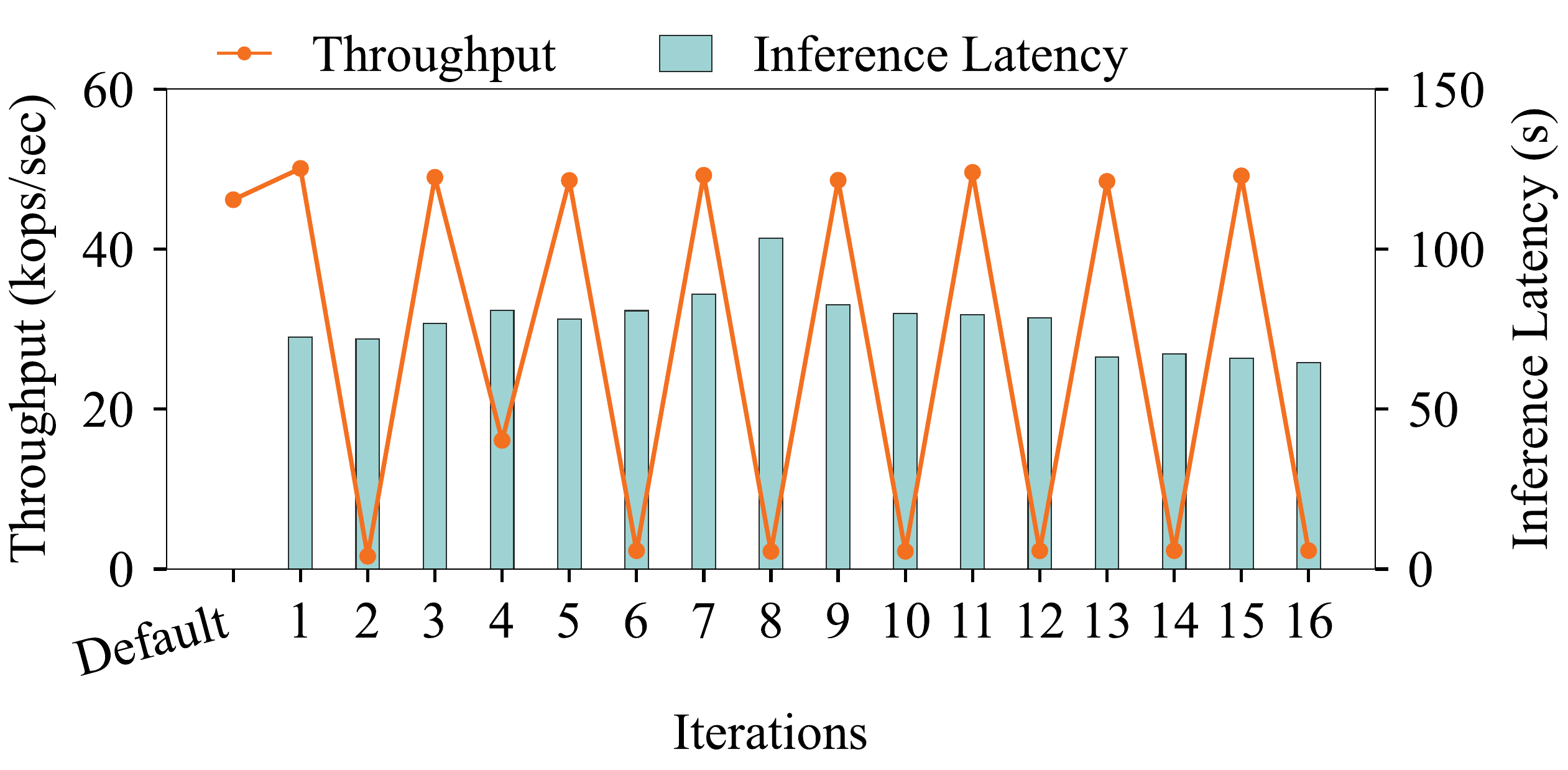}\\
    (d) \texttt{mix (1000s duration)}}
    \caption{Inference latency during each iteration. Large-scale LLM (i.e., DeepSeek-V3) tuning incurs high latency per iteration, making it unable to react to fast workload changes. (Finding \#1)}
    \label{fig:largeModel}
\end{figure}

\textbf{Finding \#1: The high inference latency of large‑scale LLMs makes them unsuitable for real‑time parameter tuning, which requires rapid adaptation to dynamic workloads.}
As illustrated in Figure \ref{fig:largeModel}, each tuning iteration that relies on a cloud‑hosted LLM (e.g., 200B+ parameters) incurs a latency of up to hundreds of seconds under both write‑intensive (e.g., \texttt{fillrandom}) and read‑intensive (e.g., \texttt{readrandom}) workloads.
This delay is primarily attributed to three factors: (1) network round‑trip time between the storage system and the cloud API endpoint, (2) queuing delays in shared cloud inference services, and (3) the intrinsic computational overhead of processing large‑scale models.
Such latency not only prevents timely parameter updates but also introduces a significant mismatch between the tuning cycle and the rapidly varying I/O patterns characteristic of modern database workloads.
Moreover, the per‑invocation API cost of commercial cloud LLMs renders continuous, fine‑grained tuning economically impractical for long‑running storage systems. However, the tuning efficiency is disproportionate to the substantial inference overhead it incurs. Performance varies significantly across workloads: while write-intensive scenarios (Figures \ref{fig:largeModel}a and \ref{fig:largeModel}c) exhibit a modest improvement in throughput compared to the default configuration, read-intensive and mixed workloads (Figures \ref{fig:largeModel}b and \ref{fig:largeModel}d) suffer from severe instability. 
Root cause analysis of tuning log reveals a strategic misjudgment by DeepSeek-V3 \cite{DeepSeek}: in an attempt to aggressively maximize read throughput, the model repeatedly enabled \texttt{cache\_index\_and\_filter\_blocks=true}.
Critically, it failed to proportionally increase the \texttt{block\_cache\_size}, leading to severe cache thrashing and performance fluctuation.
In conclusion, while large‑scale LLMs possess strong reasoning capabilities, their operational characteristics are incompatible with the low‑latency, high‑frequency decision‑making required for effective real‑time compaction tuning.

\begin{figure}[htbp]
    \centering
    \parbox{0.48\columnwidth}{\centering\includegraphics[width=0.96\linewidth]{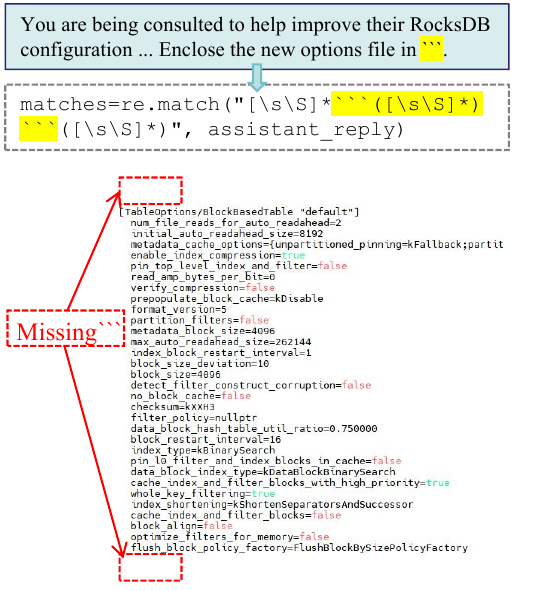}\\(a) Instruction Format Error}\vspace{0.1in}
    \parbox{0.48\columnwidth}{\centering\includegraphics[width=0.96\linewidth]{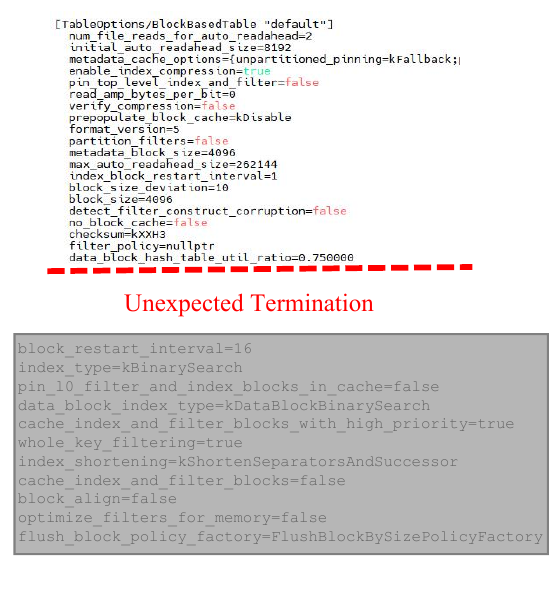}\\(b) Inference Interrupted}\vspace{0.1in}
    \parbox{0.48\columnwidth}{\centering\includegraphics[width=0.94\linewidth]{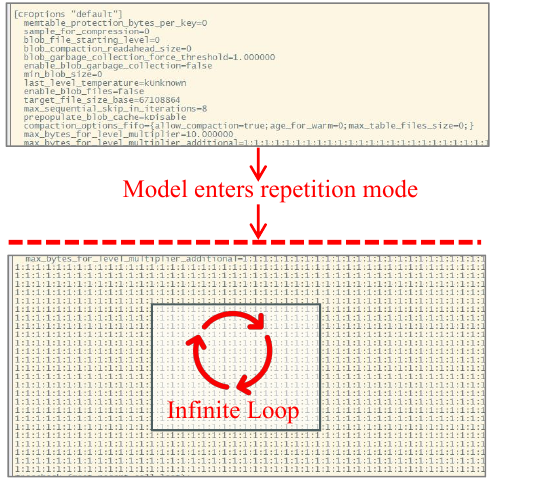}\\(c) Repeated Output}
    \parbox{0.48\columnwidth}{\centering\includegraphics[width=0.94\linewidth]{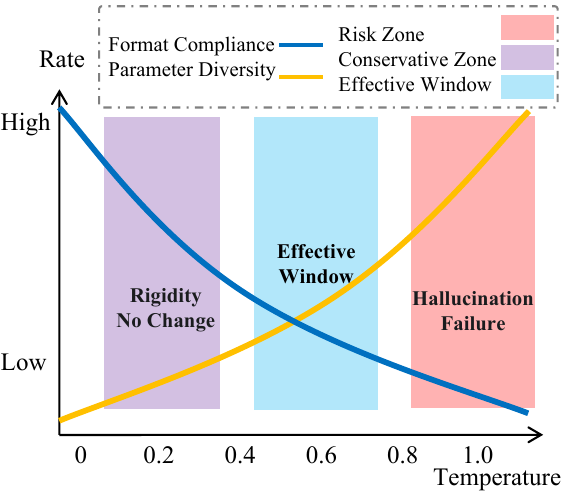}\\(d) Temperature Trade-off}
    \caption{Inference failures of 1B-parameter LLMs in LSM-tree tuning. We categorize failures into four cases:
    (a) Instruction format error, where the LLM output violates formatting constraints; 
    (b) Inference interrupted due to context limits or false stop tokens;
    (c) Repeated output, characterized by infinite generative loops; and 
    (d) Unstable exploration-stability trade-offs, showing the difficulty of finding an effective temperature window that balances format compliance and parameter diversity. (Take Pangu-1B \cite{arXiv2023pangupi} and Pangu-7B \cite{arXiv2025pangu7B} as an example. (Finding \#2)}
    \label{fig:smallModel}
\end{figure}

\textbf{Finding \#2: Limitations of the small-scale LLM in inference reasoning.}
The inference process of the small-scale LLM (typically \textless10B parameters) exhibits several critical limitations when applied to complex system optimization tasks like LSM-tree tuning.
To study the effectiveness of small models, we construct experiments on Pangu-1B \cite{arXiv2023pangupi} and Pangu-7B \cite{arXiv2025pangu7B} that are deployed at a local server.
Figure \ref{fig:smallModel} illustrates four limitations of the small-scale LLM used in the LSM-tree parameter tuning.
First, the limited parameter capacity and computational capability of small models restrict contextual reasoning, resulting in the inability to capture the intricate causal relationships between multiple tuning parameters.
Even when prompts explicitly specify the output format, small-scale LLMs often fail to adhere to such constraints strictly (Figure \ref{fig:smallModel}(a)).
This results in the failure to parse a valid optional file correctly, ultimately causing runtime errors in RocksDB.
Second, due to limited context windows or mispredicted end-of-sequence tokens, the small-scale LLM often fails to complete the generation of long configuration files (Figure \ref{fig:smallModel}(b)).
This results in truncated outputs where critical tail-end parameters are missing, leading to valid but incomplete configurations that fail during runtime.
Third, when the model is uncertain, or the context is too long (Figure \ref{fig:smallModel}(c)), it may fall into a degenerative state, repeating the same token sequence indefinitely (e.g., 1:1:1...).
The model may invent configuration parameters that do not exist in the optional file, causing the system to crash upon deployment.
Finally, the most significant challenge lies in balancing parameter exploration with format stability (Figure \ref{fig:smallModel}(d)).
The rigidity problem (T→0): Automated tuning requires the model to propose novel parameter values to optimize performance. 
However, at low temperatures, the Pangu-7B model exhibits extreme rigidity, often reproducing default values or copying the input context verbatim without performing any optimization.
The collapse problem (T>0.7): Increasing the temperature to encourage exploration introduces stochasticity that disproportionately affects syntax coherence.
Unlike large-scale LLMs, which maintain formatting even at higher temperatures, smaller models rapidly degenerate into hallucinations or invalid formats, rendering the output unparsable.
The effective window where the model is both creative enough to optimize and stable enough to parse is vanishingly narrow. 
This makes identifying and sustaining a reliable, effective zone a formidable challenge in practical tuning pipelines.
\textit{Unless otherwise specified, all subsequent references to Pangu in this work denote the 7B‑parameter version.}

\begin{figure}[htbp]
\centering
\includegraphics[width=0.9\linewidth]{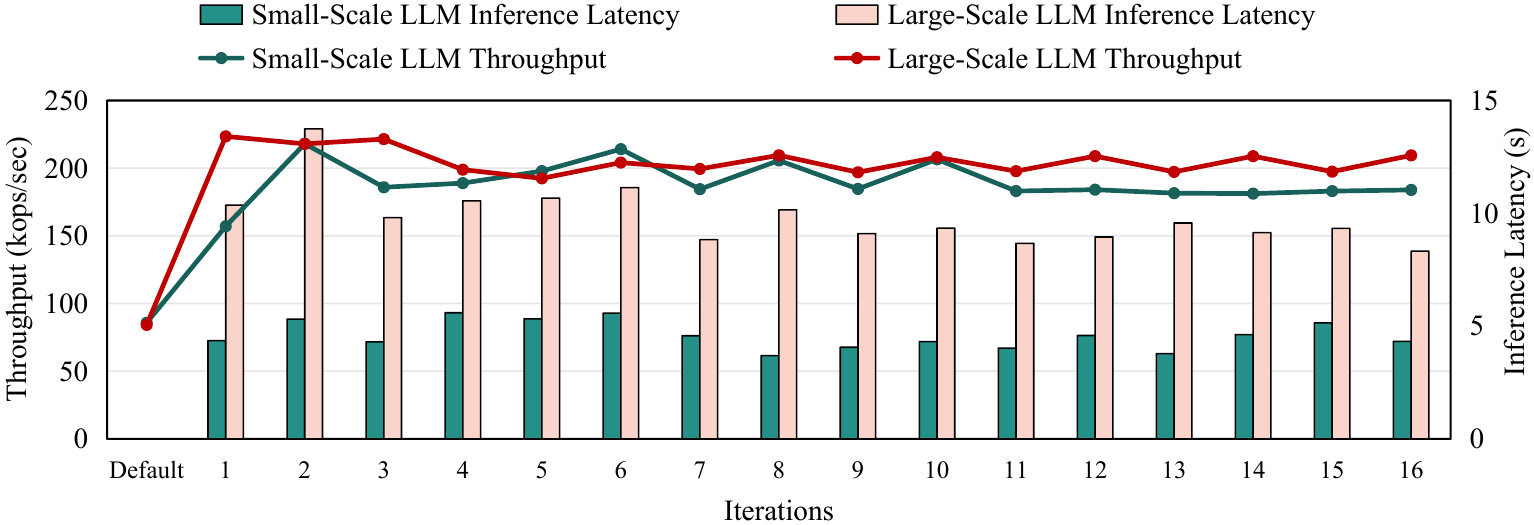}
\caption{Performance during each iteration. Large-scale LLM (i.e., DeepSeek-V3) inference results in higher inference latency than that of small-scale LLM (i.e., Pangu-7B). (Finding \#3)}
\label{fig:inferenceLat}
\end{figure}

\textbf{Finding \#3: Model scale correlates with reasoning efficiency and inference latency. Small-scale LLMs can achieve competent reasoning within a constrained parameter subspace.}
To investigate the potential of LSM-tree parameter tuning via large-scale LLMs and small-scale LLMs, we established a closed-loop iterative framework.
In this setup, the model receives the current throughput, CPU and memory usage, and the active configuration file as context to generate an optimized configuration for the next iteration.
We construct the experiments using db\_bench under a write-dominated workload \texttt{fillrandom} (1000s random writes with database reset).
To ensure experimental consistency, the system cache was cleared before each iteration.
Addressing the susceptibility of the small-scale LLM to hallucinations as discussed in Finding \#2, we simplify the tuning process that reduces the number of tuned parameters for the small-scale LLM (i.e., Pangu-7B \cite{arXiv2025pangu7B}).
Figure \ref{fig:inferenceLat} illustrates the inference latency and throughput.
There is a strong positive correlation between model parameter count and the quality of tuning decisions.
For write-intensive workload like \texttt{fillrandom}, both models rapidly improved upon the default configuration, reaching near-optimal throughput after iteration 2.
The small-scale LLM also achieves remarkable stability.
In addition, the inference latency shows that Pangu-7B achieved a lower inference time compared to DeepSeek-V3.

\textbf{Summary.} These findings motivate us to consider whether the small-scale LLM (i.e., Qwen3-8B \cite{arXiv2025Qwen3}, Pangu-7B \cite{arXiv2025pangu7B}) can achieve real-time tuning with high compaction performance while maintaining low inference latency and overhead.
Specifically, we seek to answer the following questions in our future work:
\begin{itemize}
    \item[\texttt{Q1:}] How can we enable sub-10B parameter lightweight LLMs to generate accurate LSM-tree parameter tuning?
    \item[\texttt{Q2:}] How can workload characteristics guide the reduction of inference overhead (i.e., iteration counts) while preserving real-time tuning quality?
    \item[\texttt{Q3:}] How can the trade-off between model size, inference latency, and tuning accuracy be quantitatively characterized to guide lightweight LLM selection for edge-side real-time inference scenarios?
\end{itemize}

\section{Related Work}\label{sec:relatedwork}
\textbf{Compaction Optimization.}
Compaction is a determinant of performance in LSM-based storage systems, and its optimization has been extensively studied from multiple angles.
Early studies focused on improving compaction policies to reduce write amplification and improve throughput.
TRIAD \cite{ATC2017TRIAD} introduced a tiered compaction strategy that separates hot and cold data, significantly reducing write amplification for update-intensive workloads.
SILK \cite{ATC2019SILK} coordinated compaction scheduling with the underlying file system to minimize I/O interference between foreground operations and background maintenance.
More recent work has explored learned compaction strategies: Dremel \cite{2022Dremel} employs machine learning to predict the cost-benefit ratio of potential compaction tasks, enabling more informed selection of which SSTables to merge.
VigilKV \cite{ATC2022vigilkv} implemented fine-grained I/O scheduling that prioritizes compaction I/O based on its impact on read tail latency.
At the architectural level, vLSM \cite{2024vLSM} proposed a virtual compaction layer that allows multiple compaction policies to coexist and be switched dynamically at runtime, providing adaptability to changing workload patterns.
Our work complements this line of research by enabling real-time, adaptive parameter adjustment specifically for compaction operations, allowing the system to dynamically reconfigure its tuning parameters in response to observed workload behavior.

\textbf{Parameter Tuning.}
Several works leverage the automatic parameter tuning to optimize the performance of the LSM-tree. 
Search-based optimization techniques, such as Bayesian optimization \cite{EuroMLSys2021Bayesian}, and reinforcement-learning \cite{SIGMOD2023learning}, treat the parameter space as a black box and iteratively sample configurations to find optimal settings.
These methods can achieve good results but suffer from high sample complexity, each configuration evaluation requires running the database under a representative workload, making them too slow for real-time adaptation. 
More recently, learning-based methods have gained prominence.
QTune \cite{VLDB2019QTune} uses deep reinforcement learning to tune database parameters, while ResTune \cite{SIGMOD2021Restune} incorporates residual learning to adapt to previously unseen workloads.
The most directly relevant to our work are language-model-based approaches: ELMo-Tune \cite{HotStorage2024ELMo-Tune} and its enhanced version \cite{ELMo-Tune-V2} leverage large language models trained on expert tuning traces to generate configuration advice.
However, these systems assume a cloud-based deployment where the LLM runs on powerful remote servers, incurring significant network and inference latency that makes them unsuitable for real-time, closed-loop tuning.
Our contribution shifts the tuning paradigm toward the storage edge by applying a small-scale LLM specifically to compaction-focused parameter optimization, thereby retaining necessary reasoning capability while achieving the low latency required for real-time compaction management.

\textbf{On-Device LLM.}
The deployment of LLMs on edge devices has transitioned from large-scale general reasoning to domain-specific system optimization. 
Our work aims at the intersection of LSM-tree parameter tuning of small-scale LLMs. 
The research landscape for on-device LLMs has primarily focused on overcoming resource constraints to preserve general intelligence. 
Prior works such as Cambricon-LLM \cite{yu2024cambricon} and Lincoln \cite{sun2025lincoln} have explored chiplet-based hybrid architectures and compute-enabled Flash memory to bypass I/O bottlenecks for models exceeding 50B parameters.
Similarly, LLM in a Flash \cite{alizadeh2024llm} presented a storage-to-DRAM demand loading strategy, leveraging activation sparsity and row-column bundling to run models twice the size of available DRAM on commodity hardware.
More recently, CLONE \cite{tian2025clone} utilized intelligent resource management and system-level scaling to balance inference latency with the limited DRAM of edge devices. However, while these advancements demonstrate that executing large-scale models locally is technically feasible, the associated power consumption and computational overhead remain disproportionately high for continuous, background system tasks. In energy-sensitive edge environments, the brute-force execution of massive models is often unsustainable, catalyzing a strategic shift toward specialized SLMs that achieve domain-specific expertise through fine-tuning. We argue that the true potential of on-device LLMs lies beyond conversational tasks, particularly in autonomous system optimization such as LSM-tree parameter tuning.
Unlike generic reasoning, database tuning requires a system-aware agent that can interpret complex storage metrics (e.g., write amplification and compaction latency) with surgical precision and real-time responsiveness under strict energy budgets. 
In this context, we study that leveraging PEFT techniques like LoRA \cite{hu2022lora} to create specialized, sub-billion parameter system experts represents a significant breakthrough. This approach would allow for a semantic understanding of low-level database interactions, enabling autonomous configurations that are far more responsive and efficient than traditional black-box methods.
This provides the potential to redefine how storage engines adapt to dynamic edge workloads, providing a novel frontier that general-purpose frameworks have yet to explore.

\section{Conclusion}\label{sec:conclusion}
In this paper, we present the empirical study of on-device LLM inference for real-time LSM-tree compaction optimization. 
Our results reveal a trade-off: large models deliver superior tuning accuracy but exceed latency budgets, while small models meet real-time constraints yet suffer from conservative parameter selection, recurring reasoning errors, and high output inconsistency.
However, when the input size is reduced, device-side LLM can effectively optimize the compaction-related parameters of the LSM-tree.
In our future work, we will design a real-time LSM-tree compaction optimization via on-device LLM.

\bibliographystyle{ACM-Reference-Format}
\bibliography{reference}

\end{document}